\documentclass[11pt,a4paper,twocolumn]{article}
\pdfoutput=1
\usepackage{graphicx}
\usepackage[utf8]{inputenc}     
\usepackage[T1]{fontenc}     
\usepackage{mathptmx}        
\usepackage{amsmath,amssymb} 
\usepackage[round]{natbib}   
\usepackage{doi}             
\usepackage{booktabs}        
\usepackage{hyperref}
\usepackage{color}
\usepackage[labelfont=bf, labelsep=period, font=small]{caption}  
\usepackage{authblk}         
\usepackage{courier}         
\usepackage{listings}        
\usepackage[bw]{dtsyntax}    

\usepackage{epsfig}
\usepackage{subfigure}
\usepackage{calc}
\usepackage{amssymb}
\usepackage{amstext}
\usepackage{amsmath}
\usepackage{amsthm}
\usepackage{pslatex}

\usepackage{float}
\usepackage[usenames,dvipsnames]{xcolor}
\usepackage{tcolorbox}
\usepackage{tabularx}
\usepackage{array}
\usepackage{colortbl}

\usepackage{amssymb}
\usepackage{amsmath,amssymb}
\usepackage[ruled,vlined]{algorithm2e}
\usepackage{amssymb}
\usepackage{amsmath}

\usepackage{geometry}            
\hypersetup{
	pdftitle={Paper SIMS 56 Conference},
	pdfauthor={Author Name1, Author Name2},
	hidelinks,
	pdfpagelayout=SinglePage,
	pdfcreator=pdflatex,
	pdfproducer=pdflatex}

\renewcommand{\normalsize}{\fontsize{10.5pt}{12.4pt}\selectfont}
\renewcommand{\small}{\fontsize{9.5pt}{11.2pt}\selectfont}
\renewcommand{\footnotesize}{\fontsize{8.5pt}{10.0pt}\selectfont}

\begin{document}
\thispagestyle{empty}

\title{\textbf{Robust Simulation for Hybrid Systems: \\ \textit{Chattering Bath Avoidance}}}
\renewcommand\Authfont{\large}        
\renewcommand\Affilfont{\normalsize}       
\renewcommand\Authsep{\quad}                     
\renewcommand\Authand{\quad}                     
\renewcommand\Authands{\quad}                    
\author[1]{Ayman ALJARBOUH}
\author[1]{Benoit CAILLAUD}
\affil[1]{INRIA/IRISA Rennes, Campus de Beaulieu, 35042 Rennes Cedex, France, {\small\texttt{\{ayman.aljarbouh,benoit.caillaud\}@inria.fr}}}


\date{} 
\maketitle\thispagestyle{empty} 
\abstract{
The sliding mode approach is recognized as an efficient tool for treating the chattering behavior in hybrid systems. However, the amplitude of chattering, by its nature, is proportional to magnitude of discontinuous control. A possible scenario is that the solution trajectories may successively enter and exit as well as slide on switching manifolds of different dimensions. Naturally, this arises in dynamical systems and control applications whenever there are multiple discontinuous control variables. The main contribution of this paper is to provide a robust computational framework for the most general way to extend a flow map on the intersection of $p$ intersected $(n-1)$-dimensional switching manifolds in at least $p$ dimensions. We explore a new formulation to which we can define unique solutions for such particular behavior in hybrid systems and investigate its efficient computation/simulation. We illustrate the concepts with examples throughout the paper.}


\section{Introduction}

\noindent Hybrid systems are heterogeneous dynamical systems exhibiting both continuous and discrete dynamics. Systems of this type arise naturally in control systems where the value of a control variable may jump or whenever the laws of physics are dicontinuous. They are common in a wide variety of engineering applications dealing with multi-modal systems such like mechanical systems with impact phenomena, robotics, mechatronics,  automotive industry, power systems, process control, as well as embedded computation where program codes interact with the physical world \citep{Zhang,CAI}. In such dynamical systems, the state variables  are capable of evolving continuously (flowing) and/or evolving discontinuously (jumping) \citep{CAI}. The dynamical behavior characterized by interacting continuous variables and discrete switching logics can be captured by a mathematical framework given in terms of time-driven continuous variable dynamics (usually described by differential or difference equations) and event-driven discrete logic dynamics (whose evolutions depend on \textquotedblleft if-then-else \textquotedblright \ type of rules and may be described by automata) \citep{Antsaklis}. \\
Such interaction of continuous-time and discrete-time dynamics may lead to an infinitely fast and continuous switching between several dynamics, such behavior is called a \textit{chattering} behavior \citep{Mosterman}. This often occurs in the optimal control of continuous and hybrid systems. Chattering executions can be defined as solutions to the hybrid system having infinitely many discrete transitions in finite time.
Such behavior can be intuitively thought of as \textit{involving infinitely fast switching among different control actions or modes of operation}. Similar behavior appears in variable structure control systems and in relay control systems \citep{relay}. The numerical simulation of hybrid systems exhibiting a chattering behavior appears to come to a halt, since infinitely many transitions would need to be simulated. This requires high computational costs as small step-sizes are required to maintain the numerical precision. Chattering behavior has to be treated in an appropriate way to ensure that the numerical integration progress in a reasonable time. This has been investigated by means of different methods. In general, one may prevent the chattering in hybrid systems by inducing a smooth sliding motion on the switching manifold on which the chattering occurs \citep{diBernardo,Weiss,Guadria}. While the infinitely fast switching between modes occurs, a smooth motion along the switching surface may emerge while eliminating the fast chattering. Differential inclusions (DIs) of the Filippov type (the so-called equivalent dynamics) can be used to define a regularized solution trajectory in a small neighborhood of the switching manifold on which the chattering occurs so that the average velocity of sliding on the surface can be determined \citep{Fillipov,Martin}. Another approach (the so-called equivalent control) was proposed by Utkin \citep{Utkin}. However, the computation of the equivalent dynamics turns of to be difficult whenever the systems chatters between more than two dynamics. One of the main properties of chattering behavior is that the solution trajectory may successively enter and exit as well as slide on switching manifolds of different dimensions. Problems with this type of sliding arise naturally in dynamical systems and control applications whenever there are multiple discontinuous control variables. Such multiple sliding behavior may be the origin of non-uniqueness of solution as an inescapable and important property of heterogeneous dynamical systems with discontinuous control variables. \\
The main contribution of this paper is to provide an adequate technique to detect chattering \textquotedblleft on the fly\textquotedblright \ in real-time simulation and compute its regularization beyond the infinitely fast mode switching. In particular, we present a robust and computationally efficient framework for the most general case of $p$ intersected switching manifolds, as well as the extent of how unique solutions in such cases can be defined. A hierarchical application of convex combinations to form a differential inclusion on the intersection is employed. We provide a novel computational framework which treats  non-smoothness in the trajectory of the state variables during the regularization of chattering by a smooth correction after each integration time-step. The paper is organized as follows$:$ Section 2 provides preliminaries on the modeling of hybrid systems. Section 3 deals with the chattering behavior as well as the chattering on switching intersection. Section 4 presents a numerical approach to integrate the system dynamics while eliminating the fast switching. Finally, simulation results and conclusion of the study are given in Sections 5 and 6, respectively.
\section{Modeling of Hybrid Systems}
\textbf{Definition 1$:$ (Hybrid Automaton)} \\
A hybrid automaton $HA$ is a collection $HA=(Q,X,Init, f, I, E,G,R)$, wehre
\begin {enumerate}
\item $Q$, a finite set of discrete states $q\in Q$.
\item $X$, a finite set of continuous states variables $x_i\in X$.
\item $Init\subseteq Q\times X$, is a finite set of initial states.
\item $f : Q \times \mathbb{R}^n \to \mathbb{R}^n$, is a flow map, which describes, through a differential equation, the continuous evolution of the continuous state variable $x$.
\item $I$: $Q\rightrightarrows\mathbb{R}^n$, assigns to each location $q\in Q$ an invariant map,  which describes the conditions that the continuous state $x$ has to satisfy at this mode.
\item $E\subset Q \times Q$, is a collection of discrete transitions, which identifies the pairs ($q$, $q^\prime$) such that a transition from the mode $q$ to the mode $q^\prime$ is possible.
\item $G : E \rightrightarrows \mathbb{R}^n$, assign to each $e=(q, q^\prime )\in E$ a guard to which the continuous state $x$ must belong so that a transition from $q$ to $q^\prime$ can occur.
\item $R : E \times \mathbb{R}^n \to \mathbb{R}^n$ assign to each $e=(q, q^\prime )\in E$ and $x\in X$ a reset map $R(e,x)$, which describes, for each $e=(q, q^\prime )\in E$, the value to which the continuous state $x\in\mathbb{R}^n$ is set during a transition from mode $q$ to mode $q^\prime$.
\end {enumerate}
\textbf{Definition 2$:$ (The dynamics of a Hybrid Automaton)} \\
As a hybrid system with explicitly shown modes, the continuous behavior in the hybrid automaton is modeled by a flow map $\dot{x}=f_{q}(x)$, while the discrete behavior is modeled by a jump map $\phi_{q}(x)$ \citep{CAI}. The conditions that permit flows and/or jumps are given by the flow set $C$ and the jump set $D$, respectively, subsets of the state space. The dynamics of a hybrid automaton is given by
\begin{equation}
\mathcal{H}:\ \left \lbrace \begin{array}{ccc}
\ \ \ \dot{x} \ \ \ \ = \ f_q(x) \ \ \ \ x\in C_q\\
\left [ \begin{array}{ccc}
x^+\\
q^+\end{array} \right ] \in  \ \phi_q(x) \ \ x\in D_q\end{array} \right \rbrace
\label{simpleequation}\hspace{0.7cm}
\end{equation}
where
\begin{equation}
D_q=\bigcup_{e=(q,q^\prime)\in E}G(e); \ \  \ \ \ \ C_q=I_q;
\label{simpleequation}\hspace{1.3cm}
\end{equation}
\begin{equation}
\phi_q(x)=\bigcup_{e=(q,q^\prime);\\ x\in G(e)}\left [ \begin{array}{ccc}
R(e,x)\\
q^\prime\end{array} \right ];
\label{simpleequation}\hspace{1.1cm}
\end{equation}
\begin{equation}
C=\bigcup_{q\in Q}(C_q\times \{q\}); \ \ \ \ D=\bigcup_{q\in Q}(D_q\times \{q\})
\label{simpleequation}
\end{equation}
The discrete transitions between modes are always guarded by  zero-crossings of the guard functions, and continuous modes are always defined by a Boolean expression over discrete variables, which are piecewise constant in continuous time, this is to make sure that a mode change can only happen on discrete transitions. In general, there are three natural choices for the semantics of a zero-crossing $up(z)$\citep{Schrammel}:
\begin{itemize}
\item “At-zero” $:$ $ z(x_{k-1},t_{k-1})\leq0 \wedge z(x_k,t_k)\geq0$
\item “Contact” $:$ $ z(x_{k-1},t_{k-1})<0 \wedge z(x_k,t_k)\geq0$
\item “Crossing” $:$ $z(x_{k-1},t_{k-1})\leq0 \wedge z(x_k,t_k)>0$
\end{itemize}
Based on non-standard analysis, by the continuity of the event function $z$ in the \textquotedblleft Crossing\textquotedblright \ semantics we can deduce that $z(t)=0$ is valid at the zero-crossing point in standard semantics$:$ by standardizing $z(t)\leq0 \ \wedge \ z(t+\partial)>0$ we get $st(z(t))=st(z(t+\partial))=0$, where $\partial$ is a non-standard infinitesimal \citep{Albert}. This gives an unambiguous meaning to hybrid systems even if they contain chattering behavior. In order to allow chattering in the standard semantics, the trajectories are allowed to actually go above zero, up to a constant $\epsilon>0$. \\ \\
\textbf{Definition 3$:$ (Lie derivatives)}\\
Assume the flow map $f_q$ is analytic in its second argument, the \textit{Lie derivatives} $\mathcal{L}^k_{f_q} g_q:\mathbb{R}^n\to\mathbb{R}^n$ of a function $g_q$, also analytic in its second argument, along $f_q$, for $k>0$, is defined by$:$
\begin{equation}
\mathcal{L}^k_{f_q} g_q(x(t))=\left(\frac{\partial \mathcal{L}^{k-1}_{f_q} g_q(x(t))}{\partial x(t)}\right)\cdot f_q(x(t))
\label{simpleequation}\hspace{0.3cm}
\end{equation}
with
\begin{equation}
\mathcal{L}^0_{f_q} g_q(x(t))=g_q(x(t))
\label{simpleequation}\hspace{3.3cm}
\end{equation}
\textbf{Definition 4$:$ (Pointwise relative degree)}\\
We define the relative degree $n_q(x) : \mathbb{R}^n\to\mathbb{N}$ by$:$
\begin{equation}
n_q(x)=k \ if \ \bigwedge_{j<k}\mathcal{L}^{j}_{f_q} g_q(x)=0\wedge \mathcal{L}^k_{f_q} g_q(x)\not=0
\label{simpleequation}
\end{equation}
\section{The Chattering Behavior}
\noindent Physically, chattering behavior occurs if equal thresholds for the transition conditions of different modes are given and the system starts to oscillate around them. A specific issue is the existence of multiple sliding modes due to intersecting switching manifolds, that is, the chattering occurs on switching manifolds of different dimension, roughly speaking, the chattering set may belong to many $\mathbb{R}^{(n-r)}$ switching manifolds of the same system, where $r\in\{1,2,...,n\}$.
To illustrate this particular case, we study, in the following, a mechanical stick-slip system with dry Coulomb friction.
\subsection{Case Study 1$:$}
The non-smooth system to be investigated (Figure 1), consists of three blocks of masses $m$, $M_1$, and $M_2$,  where only the block of mass $m$ is connected to a fixed support by a linear spring of stiffness $k$ and is under the action of a sinusoidal external force $u$ generated by an actuator $P$ $(u=A\ sin(\omega\ t+\varphi))$.
We denote $x_m$, $x_{M_1}$, and $x_{M_2}$ to the position of the small mass $m$ and the two inertial masses $M_1$ and $M_2$, respectively. We denote $\mathcal{F}_1$ to the tangential contact force on the frictional interface between the small mass $m$ and the inertial mass $M_1$, and $\mathcal{F}_2$ to the tangential contact force on the frictional interface between the small mass $m$ and the inertial mass $M_2$. The friction between the inertial mass $M_2$ and the ground is neglected. The origin of the displacements $x_m$, $x_{M_1}$, and $x_{M_2}$ is taken where the spring is un-stretched.
The system's state space representation is given by
\begin{equation}
f(x): \ \left \{ {\begin{array}{cc}
\dot x_m=v_m \ \ \ \ \ \ \ \ \ \ \ \ \ \ \ \ \ \ \  \ \ \ \ \ \ \ \ \ \ \ \ \ \\
\dot v_m=\frac{1}{m}(u-kx_m-\mathcal{F}_1-\mathcal{F}_2)\\
\dot x_{M_1}=v_{M_1} \ \ \ \ \ \ \ \ \ \ \ \ \ \ \ \ \ \ \  \ \ \ \ \ \ \ \ \ \\
\dot v_{M_1}=\frac{1}{M_1}\mathcal{F}_1 \ \ \ \ \ \ \ \ \ \ \ \ \ \ \ \ \ \ \ \ \ \ \ \ \ \\
\dot x_{M_2}=v_{M_2} \ \ \ \ \ \ \ \ \ \ \ \ \ \ \ \ \ \ \  \ \ \ \ \ \ \ \ \ \\
\dot v_{M_2}=\frac{1}{M_2}\mathcal{F}_2 \ \ \ \ \ \ \ \ \ \ \ \ \ \ \ \ \ \ \ \ \ \ \ \
\end{array} } \right \}
\label{simpleequation}
\end{equation}
where $x=[x_m \ v_m \ x_{M_1} \ v_{M_1} \ x_{M_2} \ v_{M_2}]^T$, and $v_m$, $v_{M_1}$, and $v_{M_2}$ are the velocities of the small mass $m$ and the inertial masses $M_1$ and $M_2$, respectively.
In modeling friction-induced vibration and noise problems, friction force is often treated phenomenologically, as a function of relative velocity between surfaces.
\begin{equation}
\mathcal{F}_1 =  \left \{ {\begin{array}{cc}  -F_{c_1} \ \ for \ \ v_{r_1}(t)<0\\ \color{white}-\color{black}F_{c_1} \ \ for \ \ v_{r_1}(t)>0 \end{array} } \right \}
\label{simpleequation}
\end{equation}
\begin{equation}
\mathcal{F}_2 =  \left \{ {\begin{array}{cc}  -F_{c_2} \ \ for \ \ v_{r_2}(t)<0\\ \color{white}-\color{black}F_{c_2} \ \ for \ \ v_{r_2}(t)>0 \end{array} } \right \}
\label{simpleequation}
\end{equation}
\begin{figure}[t]
\centering
\includegraphics[width=0.35 \textwidth]{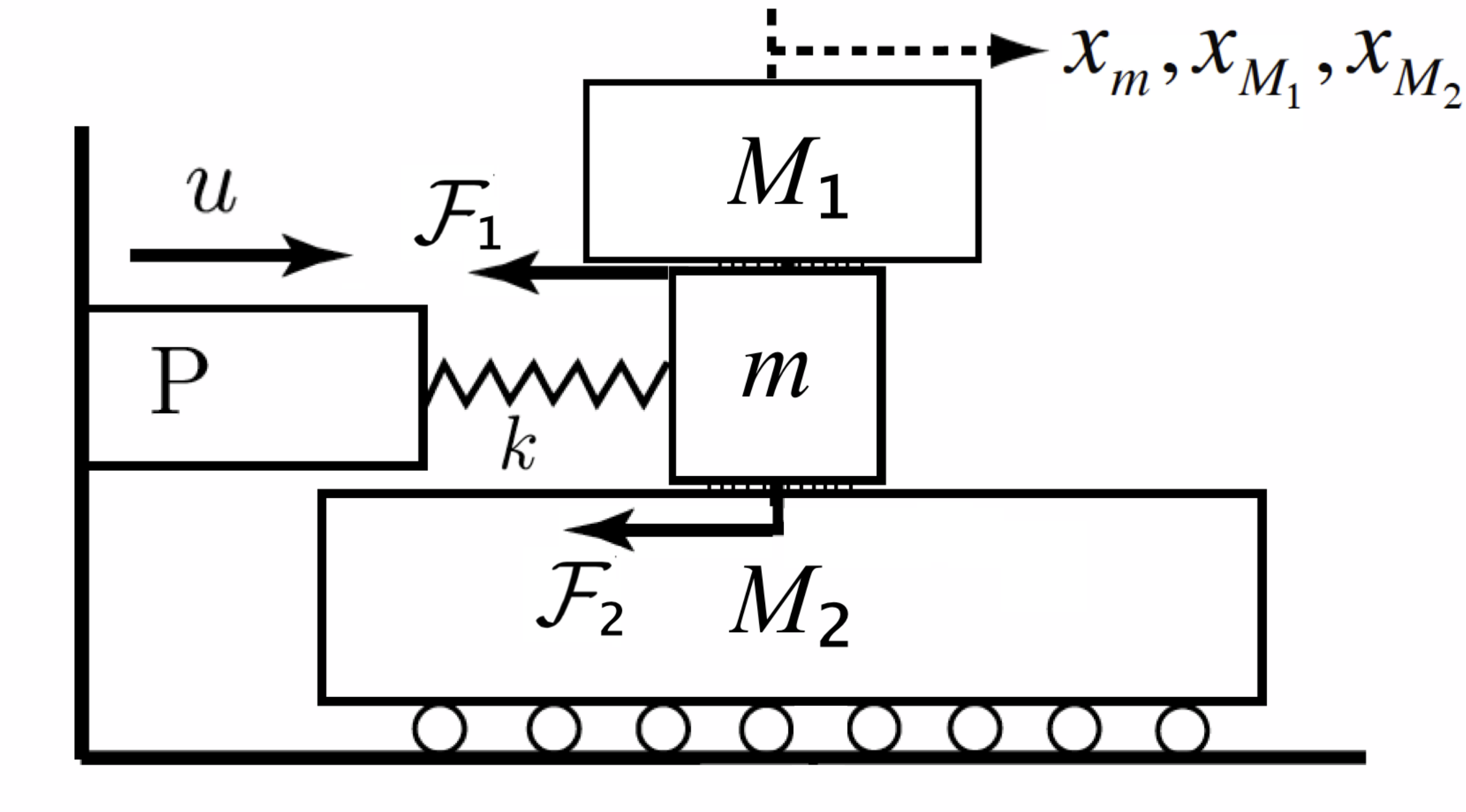}
\caption{Schematic of the studied Stick-slip system.}
\label{fig:figure1}
\end{figure}
where $F_{c_1}$, $F_{c_2}$ are the levels of the Coulomb friction, $v_{r_1}=v_m-v_{M_1}$ is the relative velocity related to the masses $m$ and $M_1$, and $v_{r_2}=v_m-v_{M_2}$ is the relative velocity related to the masses $m$ and $M_2$.
In the physical system, for the frictional interface between the small mass $m$ and the inertial mass $M_1$, as long as the force acting at the interface (call it $\rho_1$) does not exceed the Coulomb friction level $F_{c_1}$, the two masses $m$ and $M_1$ move together with $v_{r_1} = 0$. As soon as $\rho_1$ exceeds the Coulomb friction level, one mass slips over the other with $v_{r_1}\not= 0$. The slip is said to be positive ($Slip_1^+$) with a friction force $\mathcal{F}_1 = +F_{c_1}$ if $v_m > v_{M_1}$. The slip motion is said to be negative ($Slip_1^-$) with a friction force $\mathcal{F} = -F_{c_1}$ if $v_{M_1} > v_m$. The same applies for the interface between the masses $m$ and $M_2$.  \\
Obviously, we have a hybrid system with four explicitly shown modes $Q = \{q_1, q_2, q_3, q_4\}$, where the four modes $q_1 = slip_1^+slip_2^+$, $q_2 = slip_1^-slip_2^+$, $q_3 = slip_1^-slip_2^-$, and $q_4 = slip_1^+slip_2^-$ are distinguished by negative and positive relative velocities $v_{r_1}$ and $v_{r_2}$.
The hybrid automaton of this system is given, in the form (2), by
\begin{equation}
\mathcal{H}:\ \left \lbrace \begin{array}{ccc}
\ \ \ \dot{x} \ \ \ \ = \ f_q(x) \ \ x\in C_q\\
\left [ \begin{array}{ccc}
x^+\\
q^+\end{array} \right ] \in  \ \phi_q(x) \ \ x\in D_q\end{array} \right \rbrace
\label{simpleequation}
\end{equation}
\begin{equation}
f_1(x): \left \{ {\begin{array}{cc}
\dot x_m=v_m \ \ \ \ \ \ \ \ \ \ \ \ \ \ \ \ \ \ \  \ \ \ \ \ \ \ \ \ \  \ \ \ \ \  \ \ \ \ \\
\dot v_m=\frac{1}{m}\left(u-kx_m(t)-F_{c_1}-F_{c_2}\right)\\
\dot x_{M_1}=v_{M_1} \ \ \ \ \ \ \ \ \ \ \ \ \ \ \  \ \ \ \ \ \ \ \ \ \ \ \ \ \ \  \ \ \ \ \\
\dot v_{M_1}=\frac{1}{M_1}F_{c_1} \ \ \ \ \ \ \ \ \ \ \ \ \ \ \ \ \ \ \ \ \ \ \ \ \ \ \ \ \ \ \\
\dot x_{M_2}=v_{M_2} \ \ \ \ \ \ \ \ \ \ \ \ \ \ \ \  \ \ \ \ \ \ \ \ \ \ \ \ \ \  \ \ \ \ \\
\dot v_{M_2}=\frac{1}{M_2}F_{c_2} \ \ \ \ \ \ \ \ \ \ \ \ \ \ \ \ \ \ \ \ \ \ \ \ \ \ \ \ \ \
\end{array} } \right \}
\label{simpleequation}
\end{equation}
\begin{equation}
f_2(x): \left \{ {\begin{array}{cc}
\dot x_m=v_m \ \ \ \ \ \ \ \ \ \ \ \ \ \ \ \ \ \ \  \ \ \ \ \ \ \ \ \ \  \ \ \ \ \  \ \ \ \ \\
\dot v_m=\frac{1}{m}\left(u-kx_m(t)+F_{c_1}-F_{c_2}\right)\\
\dot x_{M_1}=v_{M_1} \ \ \ \ \ \ \ \ \ \ \ \ \ \ \  \ \ \ \ \ \ \ \ \ \ \ \ \ \ \  \ \ \ \ \\
\dot v_{M_1}=\frac{-1}{M_1}F_{c_1} \ \ \ \ \ \ \ \ \ \ \ \ \ \ \ \ \ \ \ \ \ \ \ \ \ \ \ \ \ \ \\
\dot x_{M_2}=v_{M_2} \ \ \ \ \ \ \ \ \ \ \ \ \ \ \ \  \ \ \ \ \ \ \ \ \ \ \ \ \ \  \ \ \ \ \\
\dot v_{M_2}=\frac{1}{M_2}F_{c_2} \ \ \ \ \ \ \ \ \ \ \ \ \ \ \ \ \ \ \ \ \ \ \ \ \ \ \ \ \ \
\end{array} } \right \}
\label{simpleequation}
\end{equation}
\begin{equation}
f_3(x): \left \{ {\begin{array}{cc}
\dot x_m=v_m \ \ \ \ \ \ \ \ \ \ \ \ \ \ \ \ \ \ \  \ \ \ \ \ \ \ \ \ \  \ \ \ \ \  \ \ \ \ \\
\dot v_m=\frac{1}{m}\left(u-kx_m(t)+F_{c_1}+F_{c_2}\right)\\
\dot x_{M_1}=v_{M_1} \ \ \ \ \ \ \ \ \ \ \ \ \ \ \  \ \ \ \ \ \ \ \ \ \ \ \ \ \ \  \ \ \ \ \\
\dot v_{M_1}=\frac{-1}{M_1}F_{c_1} \ \ \ \ \ \ \ \ \ \ \ \ \ \ \ \ \ \ \ \ \ \ \ \ \ \ \ \ \ \ \\
\dot x_{M_2}=v_{M_2} \ \ \ \ \ \ \ \ \ \ \ \ \ \ \ \  \ \ \ \ \ \ \ \ \ \ \ \ \ \  \ \ \ \ \\
\dot v_{M_2}=\frac{-1}{M_2}F_{c_2} \ \ \ \ \ \ \ \ \ \ \ \ \ \ \ \ \ \ \ \ \ \ \ \ \ \ \ \ \ \
\end{array} } \right \}
\label{simpleequation}
\end{equation}
\begin{equation}
f_4(x): \left \{ {\begin{array}{cc}
\dot x_m=v_m \ \ \ \ \ \ \ \ \ \ \ \ \ \ \ \ \ \ \  \ \ \ \ \ \ \ \ \ \  \ \ \ \ \  \ \ \ \ \\
\dot v_m=\frac{1}{m}\left(u-kx_m(t)-F_{c_1}+F_{c_2}\right)\\
\dot x_{M_1}=v_{M_1} \ \ \ \ \ \ \ \ \ \ \ \ \ \ \  \ \ \ \ \ \ \ \ \ \ \ \ \ \ \  \ \ \ \ \\
\dot v_{M_1}=\frac{1}{M_1}F_{c_1} \ \ \ \ \ \ \ \ \ \ \ \ \ \ \ \ \ \ \ \ \ \ \ \ \ \ \ \ \ \ \\
\dot x_{M_2}=v_{M_2} \ \ \ \ \ \ \ \ \ \ \ \ \ \ \ \  \ \ \ \ \ \ \ \ \ \ \ \ \ \  \ \ \ \ \\
\dot v_{M_2}=\frac{-1}{M_2}F_{c_2} \ \ \ \ \ \ \ \ \ \ \ \ \ \ \ \ \ \ \ \ \ \ \ \ \ \ \ \ \ \
\end{array} } \right \}
\label{simpleequation}
\end{equation}
\begin{equation}
\phi_1(x)=\ \left \lbrace \begin{array}{ccc}
(x,q_2) \ \ \ if \ v_{r_1}=0 \wedge v_{r_2}\geq0\\
(x,q_4) \ \ \ if \ v_{r_1}\geq0 \wedge v_{r_2}=0
\end{array} \right \rbrace
\label{simpleequation}
\end{equation}
\begin{equation}
\phi_2(x)=\ \left \lbrace \begin{array}{ccc}
(x,q_1) \ \ \ if \ v_{r_1}=0 \wedge v_{r_2}\geq0\\
(x,q_3) \ \ \ if \ v_{r_1}\leq0 \wedge v_{r_2}=0
\end{array} \right \rbrace
\label{simpleequation}
\end{equation}
\begin{equation}
\phi_3(x)=\ \left \lbrace \begin{array}{ccc}
(x,q_2) \ \ \ if \ v_{r_1}\leq0 \wedge v_{r_2}=0\\
(x,q_4) \ \ \ if \ v_{r_1}=0 \wedge v_{r_2}\leq0
\end{array} \right \rbrace
\label{simpleequation}
\end{equation}
\begin{equation}
\phi_4(x)=\ \left \lbrace \begin{array}{ccc}
(x,q_1) \ \ \ if \ v_{r_1}\geq0 \wedge v_{r_2}=0\\
(x,q_3) \ \ \ if \ v_{r_1}=0 \wedge v_{r_2}\leq0
\end{array} \right \rbrace
\label{simpleequation}
\end{equation}
\begin{equation}
D_{q_1}=G({q_1},{q_2}) \ \cup \ G({q_1},{q_4})
\label{simpleequation}\hspace{2.1cm}
\end{equation}
\[\hspace{0.7cm}=\{\left(v_{r_1}=0 \wedge v_{r_2}\geq0\right)\cup\left(v_{r_1}\geq0 \wedge v_{r_2}=0\right)\} \]
\begin{equation}
D_{q_2}=G({q_2},{q_1}) \ \cup \ G({q_2},{q_3})
\label{simpleequation}\hspace{2.1cm}
\end{equation}
\[\hspace{0.7cm}=\{\left(v_{r_1}=0 \wedge v_{r_2}\geq0\right)\cup\left(v_{r_1}\leq0 \wedge v_{r_2}=0\right)\} \]
\begin{equation}
D_{q_3}=G({q_3},{q_2}) \ \cup \ G({q_3},{q_4})
\label{simpleequation}\hspace{2.1cm}
\end{equation}
\[\hspace{0.7cm}=\{\left(v_{r_1}\leq0 \wedge v_{r_2}=0\right)\cup\left(v_{r_1}=0 \wedge v_{r_2}\leq0\right)\} \]
\begin{equation}
D_{q_4}=G({q_4},{q_1}) \ \cup \ G({q_4},{q_3})
\label{simpleequation}\hspace{2.1cm}
\end{equation}
\[\hspace{0.7cm}=\{\left(v_{r_1}\geq0 \wedge v_{r_2}=0\right)\cup\left(v_{r_1}=0 \wedge v_{r_2}\leq0\right)\} \]
Clearly, the system is discontinuous on two hyper switching manifolds $\Gamma_a$ and $\Gamma_b$ characterized as zero sets of the switching functions $\gamma_a(x)$ and $\gamma_b(x)$, respectively$:$
\begin{equation}
\Gamma_a=\{x\in\mathbb{R}^6: \gamma_a(x)\models v_{r_1}(t)=0\}
\label{simpleequation}
\end{equation}
\begin{equation}
\Gamma_b=\{x\in\mathbb{R}^6: \gamma_b(x)\models v_{r_2}(t)=0\}
\label{simpleequation}
\end{equation}
We will assume that $\frac{\partial\gamma_a(x)}{\partial x}\not=0$ for all $x\in\Gamma_a$ and $\frac{\partial\gamma_b(x)}{\partial x}\not=0$ for all $x\in\Gamma_b$ so that the normal units $\bot_a(x)$ and $\bot_b(x)$ orthogonal to the tangent planes $T_x(\Gamma_a)$ and $T_x(\Gamma_b)$, respectively, are well defined \citep{Fillipov}. Furthermore, we assume that $\bot_a(x)$ and $\bot_b(x)$ are linearly independent for all $x\in\Gamma_a\cap\Gamma_b$.
In the neighborhood of the intersection $\Delta=\Gamma_a\cap\Gamma_b$ (Figure 2), for the continuous time variables $v_m$, $v_{M_1}$, $v_{M_2}$, the $\mathbb{R}^3$ phase space consists of four open regions (the flow sets) $C_i$, for $i=1,2,3,4$, with the flow map vector fields $f_1(x)$, $f_2(x)$, $f_3(x)$, and $f_4(x)$.
The curve $\Delta$ is a $\mathbb{R}^{(n-2)}$ switching manifold results from the intersection of two transversally intersected $\mathbb{R}^{(n-1)}$ manifolds and is given by $\Delta=\Gamma_a\cap\Gamma_b=\{x\in\mathbb{R}^n: v_{r_1}=0\wedge v_{r_2}=0\}$. Clearly, the curve $\Delta$ belongs to the boundary of each one of discrete states $q_1$, $q_2$, $q_3$, and $q_4$. Furthermore, each two adjacent regions are separated by a hyper switching manifold defined by opposed zero crossings of the same switching function.
\begin{figure}[h]
\centering
\includegraphics[width=0.43 \textwidth]{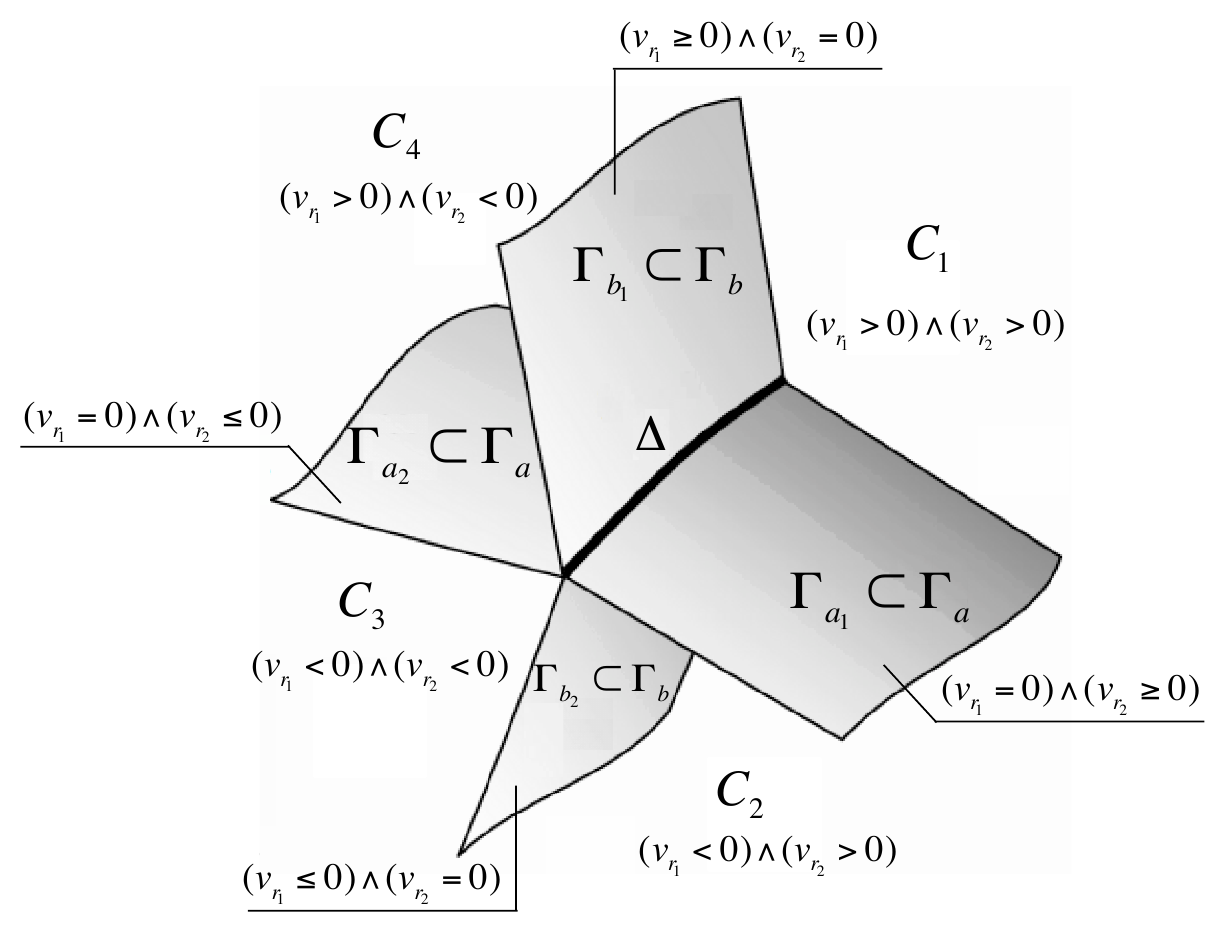}
\caption{The switching intersection set in the $\mathbb{R}^3$-dimensional phase space of the studied system.}
\label{fig:figure2}
\end{figure}
\noindent Let $\Gamma$ be the entire discontinuity region in the phase space defined as the union of the two hyper switching manifolds $\Gamma_a$ and $\Gamma_b$$:$ $\Gamma=\Gamma_a\cup\Gamma_b$. A switching between the four different flow map's vector fields $f_1$, $f_2$, $f_3$, and $f_4$ takes place in the neighborhood of the intersection $\Delta$. A trajectory that crosses $\Gamma$ transversally will switch instantaneously between the vector fields $f_q$ without any specific form of flow map vector on $\Gamma$. The only alternative is that a trajectory exhibits an attractive chattering on $\Gamma$, which yields an attractive chattering: i) between $q_1$ and $q_2$ on $\Gamma_{a_1}\in\mathbb{R}^{(n-1)}$, ii) between $q_3$ and $q_4$ on $\Gamma_{a_2}\in\mathbb{R}^{(n-1)}$, iii) between $q_1$ and $q_4$ on $\Gamma_{b_1}\in\mathbb{R}^{(n-1)}$, iv) between $q_2$ and $q_3$ on $\Gamma_{b_2}\in\mathbb{R}^{(n-1)}$, v) as well as between the four discrete states $q_1$, $q_2$, $q_3$, and $q_4$ on $\Delta\in\mathbb{R}^{(n-2)}$. On the intersection $\Delta$, an infinite number of discrete transitions between the four modes is expected as long as for all $q\in Q$ the gradient of the continuous-time behavior according to the flow map’s vector field $f_q$ is directed into the intersection $\Delta$, that is, in the neighborhood of the intersection the gradients direct behavior towards $\Delta$ so that upon each entering to a new mode an infinitesimal step causes another mode change. When simulating this system, in general, such behavior breaks down the numerical integration of the dynamics, as it does not progress in time, but chatters between modes.
\subsection{Chattering in The Neighborhood of a Switching Intersection$:$}
We consider a hybrid automaton $\mathcal{H}$ with a finite set of discrete states $q\in Q$ with transverse invariants \citep {Morris} where the state space is split into different regions (flow sets $C_q\in\mathbb{R}^{n}$) by the intersection of $p$ transversally intersected $\mathbb{R}^{n-1}$ switching manifolds $\Gamma_j$ defined as the zeros of a set of scalar functions $\gamma_j(x)$ for $j=1,2,...,p$,
\begin{equation}
\Gamma_j=\{x\in\mathbb{R}^n: \ \gamma_j(x)=0 \ ; \ \ \ j=1,2,...,p\}
\end{equation}
The zero crossing in opposite directions defines the switching between two adjacent flow sets.
All switching functions $\gamma_j$ are assumed to be analytic in their second arguments so the normal unit vector $\bot_j$ for each one of the intersected switching manifolds $\Gamma_j$ is well defined. Moreover, the normal unit vectors are linearly independent for all the $\mathbb{R}^{(n-r)}$ intersections where $r\in\{2,3,...,n\}$. The system phase space $C\subseteq \mathbb{R}^{n}$ is partitioned into $2^p$ open convex regions (sub-domains) $C_q\in\mathbb{R}^n$, $q=1,...2^p$, and $p$ switching manifold $\gamma_j(x)\in\mathbb{R}^{n-1}$, $j=1,2,...,p$. In the convex flow sets $C_q$, the solution trajectory flow is governed by different vector fields $f_q(x(t))$. Furthermore, it is assumed that the flow maps $f_q$ are smooth in the state $x$ for any open flow set adjacent to the flow set $C_q$ and all $f_q$ can be associated to the intersection.\\
In general, the disjoint $2^p$ regions flow sets can be characterized in a systematic way by defining a sign matrix $\Psi$ of size $(p,2^p)$ as following$:$\\
Let $W^{(k)}=\left(W_{i}\right)_{1,2^k}$ be an $1 \times 2^k$ vector of values $W_{i}=1, \ \forall i$, with $k=1,...,p-1$.
\begin{enumerate}
\item The first row vector of the matrix $\Psi$: $V_1=\Psi(1,:)$ consists of $2^{p-1}$ pairs of $\left(-W^{(1)}(1,1),W^{(1)}(1,2)\right)$.
\begin{equation}
\Psi^{(1)}=\{-W^{(1)}(1,1),W^{(1)}(1,2)\}_{2^{p-1}}
   \label{simpleequation}\hspace{0.0cm}
\end{equation}
\item The second row vector $V_2=\Psi(2,:)$ consists of $2^{p-2}$ pairs of $(-W^{(1)},W^{(1)})$.
\begin{equation}
\Psi^{(2)}=\{-W^{(1)},W^{(1)}\}_{2^{p-2}}
   \label{simpleequation}  \hspace{1.7cm}
\end{equation}
\item The $j^{th}$ row vector $V_j=\Psi(j,:)$ consists of $2^{p-j}$ pairs of $(-W^{(j-1)},W^{(j-1)})$ for $j=3,...,p$.
\begin{equation}
\Psi^{(j)}=\{-W^{(j-1)},W^{(j-1)}\}_{2^{p-j}}
   \label{simpleequation} \hspace{1cm}
\end{equation}
\end{enumerate}
Each column vector defines the signs of the switching functions $\gamma_j(x)$ for a given flow set $C_q$.
We use the multi-valued function $\alpha_j(x)$ such that the convex set is given for all $\Gamma_j|_{j=1,2,...,p}$ and $C_i|_{i=1,2,...,2^p}$ by$:$
\begin{equation}
\dot x\in\sum_{i=1}^{2^p}\left((\prod_{j=1}^p\frac{1+2\Psi_{j,i}\cdot\alpha_j(x)-\Psi_{j,i}}{2})\cdot f_i(x)\right)
   \label{simpleequation}
\end{equation}
where
\begin{equation}
\alpha_j(x)=\left \{ \begin{array}{ccc}
\color{white}(0, \color{black}1\color{white})\color{black} \ \ \ \ \ for \ \ \gamma_j(x)>0\\
\left[0, 1\right] \ \ \ \ \ for \ \ \gamma_j(x)=0\\
\color{white}(\color{black}0\color{white}, 1)\color{black} \ \ \ \ \ for \ \ \gamma_j(x)<0
\end{array} \right\}
   \label{simpleequation}
\end{equation}
\begin{equation}
\sum_{i=1}^{2^p}\left(\prod_{j=1}^p \frac{1+2\Psi_{j,i}\cdot\alpha_j-\Psi_{j,i}}{2}\right)=1
   \label{simpleequation} \hspace{0.5cm}
\end{equation}
Equation (31) yields in
\begin{equation}
\dot x\in(1-\alpha_j)\cdot \sum_{i=1}^{2^p}\left( R_1 \cdot f_i(x)\right)+\alpha_j\cdot \sum_{i=1}^{2^p}\left( R_2 \cdot f_i(x)\right)
   \label{simpleequation}
\end{equation}
\begin{equation}
R_1=\prod_{k=1;k\not=j;\Psi_{k,i}=-1}^p\left(\frac{1+2\Psi_{k,i}\cdot\alpha_k-\Psi_{k,i}}{2}\right)
   \label{simpleequation} \hspace{-0.4cm}
\end{equation}
\begin{equation}
R_2=\prod_{k=1;k\not=j;\Psi_{k,i}=1}^p\left(\frac{1+2\Psi_{k,i}\cdot\alpha_k-\Psi_{k,i}}{2}\right)
   \label{simpleequation} \hspace{-0.2cm}
\end{equation}
We then define a matrix $F$ of the normal projections $f_i^{\bot_j}(x)$ for $j=1,2,...,p$ and $i=1,2,...,2^p$ as
\begin{equation}
F=\left \{ \begin{array}{ccc}
f_1^{\bot_1}(x) \ \ f_1^{\bot_2}(x) \ \ \cdots \ \ f_1^{\bot_p}(x)\\
f_2^{\bot_1}(x) \ \ f_2^{\bot_2}(x) \ \ \cdots \ \ f_2^{\bot_p}(x)\\
\vdots \hspace{1.0cm} \vdots   \hspace{1.0cm}  \ddots \hspace{1.0cm} \vdots  \\
\vdots \hspace{1.0cm} \vdots   \hspace{1.0cm}  \ddots \hspace{1.0cm} \vdots  \\
f_{2^p}^{\bot_1}(x) \ \ f_{2^p}^{\bot_2}(x) \ \ \cdots \ \ f_{2^p}^{\bot_p}(x)
\end{array} \right\}
\label{simpleequation}\hspace{0.3cm}
\end{equation}
where
\begin{equation}
f_i^{\bot_j}(x)=\mathcal{L}_{f_i} \gamma_j(x)=\left(\frac{\partial \gamma_j(x)}{\partial x}\right)^T\cdot f_i (x)
   \label{simpleequation}\hspace{0.3cm}
\end{equation}
In agreement with the sign matrix $\Psi$, the attractive chattering on any $\mathbb{R}^{(n-r)}$ switching manifold  for $r=2,3,...,n$ can be easily observed by checking the signs of the matrices $F$ and $\Psi$. \\ \\
\textbf{Lemma 1$:$}\\
\textit{The sufficient condition for having an attractive chattering on any switching intersection in the system's state space requires a nodal attractivity towards the intersection itself, for all the flow maps $f_i$ in the $\mathbb{R}^{n}$ regions $C_i$ associated to this intersection. That is, the following constraint should be satisfied }
\begin{equation}
\forall i,j: \ \ sgn(f_i^{\bot_j}(x))=-sgn(\Psi_{j,i})
   \label{simpleequation}\hspace{1.0cm}
\end{equation}
To keep the solution trajectory in a sliding motion on the intersection as long as the attractive chattering condition is satisfied we impose
\[\forall j=1,2,...p: \hspace{4.0cm} \]
\begin{equation}
\sum_{i=1}^{2^p}(\prod_{j=1}^p\frac{1+2\Psi_{j,i}\cdot\alpha_j-\Psi_{j,i}}{2})\cdot f_i^{\bot_j}(x)=0
   \label{simpleequation}\hspace{-0.1cm}
\end{equation}
so that
\begin{equation}
\alpha_j=\frac{W_1}{W_1-W_2}
\label{simpleequation}\hspace{5.3cm}
\end{equation}
\begin{equation}
W_1=\sum_{i=1}^{2^p}(\prod_{k=1;k\not=j;\Psi_{k,i}=-1}^p\frac{1+2\Psi_{k,i}\cdot\alpha_k-\Psi_{k,i}}{2})\cdot f_i^{\bot_j}
   \label{simpleequation}
\end{equation}
\begin{equation}
W_2=\sum_{i=1}^{2^p}(\prod_{k=1;k\not=j;\Psi_{k,i}=1}^p\frac{1+2\Psi_{k,i}\cdot\alpha_k-\Psi_{k,i}}{2})\cdot f_i^{\bot_j}
   \label{simpleequation}
\end{equation}
For all $\alpha_k\in(0,1)$, the product term in (42) (respectively (43)) takes always a value in (0,1) since it is always a product of $(1-\alpha_k)$ (respectively $\alpha_k$). It holds always that $W_1>0\wedge W_2<0$ as long as an attractive chattering takes place at $x\in \left( \bigcup_{j=1}^p\Gamma_j \right)\cap C$, where $C$ is the entire flow set in the system phase space.\\ This gives us a hypercube convex hull of sign coordiantes ($\pm 1$, $\pm 1$, $\cdots$, $\pm 1$) with an edge of length $2$ and $\mathbb{R}^n$ volume $2^p$. Therefore, a solution to the fixed point non-linear problem (40) exists. However, the uniqueness of the solution is no guaranteed.
To deal with non-uniqueness on the intersection$-$on which the attractive chattering occurs$-$we propose to give an equivalent to the product term in (40) so that the sliding parameters are given in term of a rational function of coefficients $\kappa_i$$:$
\begin{equation}
\prod_{j=1}^p\frac{1+2\Psi_{j,i}\cdot\alpha_j-\Psi_{j,i}}{2}=\frac{\kappa_i}{\sum_{k=1}^{2^p}\kappa_k}
\label{simpleequation}\hspace{1.1cm}
\end{equation}
where
\begin{equation}
\kappa_i=\frac{\left(\prod_{l=1;l\not=i}^{2^p}(\Omega_l)\right)^{\frac{1}{{2^p}-1}}}{\left(\prod_{l=1;l\not=i}^{2^p}(\Omega_l)\right)^{\frac{1}{{2^p}-1}}-{(\Omega_i)}}
\label{simpleequation}\hspace{1.3cm}
\end{equation}
\begin{equation}
\Omega_i=\left [ (b_i)_1\ (b_i)_2 \ \cdots \ (b_i)_p \right]\cdot\left [ \begin{array}{ccc}
\mathcal{L}_{f_i} \gamma_1(x)\\
\mathcal{L}_{f_i} \gamma_2(x)\\
\vdots\\
\mathcal{L}_{f_i} \gamma_p(x)
\end{array} \right]
\end{equation}
\begin{equation}
\Omega_l=\left [ (b_l)_1\ (b_l)_2 \ \cdots \ (b_l)_p \right]\cdot\left [ \begin{array}{ccc}
\mathcal{L}_{f_l} \gamma_1(x)\\
\mathcal{L}_{f_l} \gamma_2(x)\\
\vdots\\
\mathcal{L}_{f_l} \gamma_p(x)
\end{array} \right]
\end{equation}
The vectors $b_n$ for $n=i,l$ are given as sign permutations of coordinates $[\pm 1, \pm 1, \cdots, \pm 1]^T$ under the constraint$:$
\begin{equation}
sgn(b_n)_j=\left \{ \begin{array}{ccc}
\color{white}-\color{black}sgn(\mathcal{L}_{f_n} \gamma_j(x)) \ \ for \ \ n=1\color{white},3,...,2^p\color{black}\\
-sgn(\mathcal{L}_{f_n} \gamma_j(x)) \ \ for \ \ n=2,3,...,2^p\\
\end{array} \right\}
\end{equation}
where $j=1,2,...,p$ and $p$ is the number of the intersected $\mathbb{R}^{(n-1)}$ switching manifolds.
This gives always $\Omega_1>0$ and $\Omega_n<0$ for all $n\in\{2,3,..,2^p\}$ which is exactly what we want. Another advantage of using the signs constraint in (48) is that $\kappa_j=0$ for all $j\not=i$ when $\kappa_i=1$ for a given index $i\in\{1,2,...,2^p\}$, this allows us to detect when a switching regime of different dimension has been reached by the solution trajectory, and then, to select the appropriate vector fields on this regime. Moreover, the parameter $\kappa_i$ takes always a value $0\leq\kappa_i\leq 1$ for $i=1,2,...,2^p$, yields in $\sum_{i=1}^{2^p}\left(\frac{\kappa_i}{\sum_{k=1}^{2^p}\kappa_k}\right)=1$, which is consistent with the approach of Filippov differential inclusion. \\ \\
\textbf{Case Study 1 Revisited$:$} \\
With $A=u-kx_m$, the Lie derivatives of the switching functions $\gamma_a(x)$ and $\gamma_b(x)$ along the flow map vector fields $f_1$, $f_2$, $f_3$, and $f_4$ are given by
\begin{equation}
f_1^{\bot_a}(x)=\frac{1}{m}\cdot\left(A-(\frac{m+M_1}{M_1}) \cdot F_{c_1}-F_{c_2}\right)
\end{equation}
\begin{equation}
f_1^{\bot_b}(x)=\frac{1}{m}\cdot\left(A-F_{c_1}-(\frac{m+M_2}{M_2})\cdot F_{c_2}\right)
\end{equation}
\begin{equation}
f_2^{\bot_a}(x)=\frac{1}{m}\cdot\left(A+(\frac{m+M_1}{M_1}) \cdot F_{c_1}-F_{c_2}\right)
\end{equation}
\begin{equation}
f_2^{\bot_b}(x)=\frac{1}{m}\cdot\left(A+F_{c_1}-(\frac{m+M_2}{M_2}) \cdot F_{c_2}\right)
\end{equation}
\begin{equation}
f_3^{\bot_a}(x)=\frac{1}{m}\cdot\left(A+(\frac{m+M_1}{M_1}) \cdot F_{c_1}+F_{c_2}\right)
\end{equation}
\begin{equation}
f_3^{\bot_b}(x)=\frac{1}{m}\cdot\left(A+F_{c_1}+(\frac{m+M_2}{M_2}) \cdot F_{c_2}\right)
\end{equation}
\begin{equation}
f_4^{\bot_a}(x)=\frac{1}{m}\cdot\left(A-(\frac{m+M_1}{M_1}) \cdot F_{c_1}+F_{c_2}\right)
\end{equation}
\begin{equation}
f_4^{\bot_b}(x)=\frac{1}{m}\cdot\left(A-F_{c_1}+(\frac{m+M_2}{M_2}) \cdot F_{c_2}\right)
\end{equation}.
We distinguish between the following cases$:$
\begin{itemize}
\item An infinitely fast back and forth switching between the discrete states $q_1$ and $q_2$ occurs on $\Gamma_{a_1}$ if and only if $f_1^{\bot_a}(x)<0\wedge f_2^{\bot_a}(x)>0$. This happens when $|\left(u-k x_m-F_{c_2}\right)|<(\frac{m+M_1}{M_1})\cdot F_{c_1}$. The resulting sliding vector field $f_{s_{a_1}}$ is$:$
\begin{equation}
f_{s_{a_1}}=\left \{ {\begin{array}{cc}  v_m\ \ \ \ \ \ \ \ \ \ \ \ \ \ \ \ \ \ \ \ \ \ \ \ \ \ \\\ \frac{1}{m+M_1} (u-kx_m-F_{c_2})\\ v_{M_1}\ \ \ \ \ \ \ \ \ \ \ \ \ \ \ \ \ \ \ \ \ \ \ \ \ \\ \frac{1}{m+M_1} (u-kx_m-F_{c_2})\\v_{M_2}\ \ \ \ \ \ \ \ \ \ \ \ \ \ \ \ \ \ \ \ \ \ \ \ \ \\ \frac{1}{M_2} F_{c_2}\ \ \ \ \ \ \ \ \ \ \ \ \ \ \ \ \ \ \ \ \ \ \end{array} } \right \}
\label{simpleequation}\hspace{1.0cm}
\end{equation}
\item An infinitely fast back and forth switching between the discrete states $q_3$ and $q_4$ occurs on $\Gamma_{a_2}$ if and only if $f_4^{\bot_a}(x)<0\wedge f_3^{\bot_a}(x)>0$. This happens when $|\left(u-k x_m+F_{c_2}\right)|<(\frac{m+M_1}{M_1})\cdot F_{c_1}$. The resulting sliding vector field $f_{s_{a_2}}$ is$:$
\begin{equation}
f_{s_{a_2}}=\left \{ {\begin{array}{cc}  v_m\ \ \ \ \ \ \ \ \ \ \ \ \ \ \ \ \ \ \ \ \ \ \ \ \ \ \\\ \frac{1}{m+M_1} (u-kx_m+F_{c_2})\\ v_{M_1}\ \ \ \ \ \ \ \ \ \ \ \ \ \ \ \ \ \ \ \ \ \ \ \ \ \\ \frac{1}{m+M_1} (u-kx_m+F_{c_2})\\v_{M_2}\ \ \ \ \ \ \ \ \ \ \ \ \ \ \ \ \ \ \ \ \ \ \ \ \ \\ \frac{-1}{M_2} F_{c_2}\ \ \ \ \ \ \ \ \ \ \ \ \ \ \ \ \ \ \ \ \ \ \end{array} } \right \}
\label{simpleequation}\hspace{1.0cm}
\end{equation}
\item An infinitely fast back and forth switching between the discrete states $q_1$ and $q_4$ occurs on $\Gamma_{b_1}$ if and only if $f_1^{\bot_b}(x)<0\wedge f_4^{\bot_b}(x)>0$. This happens when $|\left(u-k x_m-F_{c_1}\right)|<(\frac{m+M_2}{M_2})\cdot F_{c_2}$. The resulting sliding vector field $f_{s_{b_1}}$ is$:$
\begin{equation}
f_{s_{b_1}}=\left \{ {\begin{array}{cc}  v_m\ \ \ \ \ \ \ \ \ \ \ \ \ \ \ \ \ \ \ \ \ \ \ \ \ \ \\\ \frac{1}{m+M_2} (u-kx_m-F_{c_1})\\ v_{M_1}\ \ \ \ \ \ \ \ \ \ \ \ \ \ \ \ \ \ \ \ \ \ \ \ \ \\ \frac{1}{M_1} F_{c_1}\ \ \ \ \ \ \ \ \ \ \ \ \ \ \ \ \ \ \ \ \ \ \\v_{M_2}\ \ \ \ \ \ \ \ \ \ \ \ \ \ \ \ \ \ \ \ \ \ \ \ \ \\ \frac{1}{m+M_2} (u-kx_m-F_{c_1})\end{array} } \right \}
\label{simpleequation}\hspace{1.0cm}
\end{equation}
\item An infinitely fast back and forth switching between the discrete states $q_2$ and $q_3$ occurs on $\Gamma_{b_2}$ if and only if $f_2^{\bot_b}(x)<0\wedge f_3^{\bot_b}(x)>0$. This happens when $|\left(u-k x_m+F_{c_1}\right)|<(\frac{m+M_2}{M_2})\cdot F_{c_2}$. The resulting sliding vector field $f_{s_{b_2}}$ is$:$
\begin{equation}
f_{s_{b_2}}=\left \{ {\begin{array}{cc}  v_m\ \ \ \ \ \ \ \ \ \ \ \ \ \ \ \ \ \ \ \ \ \ \ \ \ \ \\\ \frac{1}{m+M_2} (u-kx_m+F_{c_1})\\ v_{M_1}\ \ \ \ \ \ \ \ \ \ \ \ \ \ \ \ \ \ \ \ \ \ \ \ \ \\ \frac{-1}{M_1} F_{c_1}\ \ \ \ \ \ \ \ \ \ \ \ \ \ \ \ \ \ \ \ \ \ \\v_{M_2}\ \ \ \ \ \ \ \ \ \ \ \ \ \ \ \ \ \ \ \ \ \ \ \ \ \\ \frac{1}{m+M_2} (u-kx_m+F_{c_1})\end{array} } \right \}
\label{simpleequation}\hspace{1.0cm}
\end{equation}
\item An infinitely fast switching between $q_1$, $q_2$, $q_3$, and $q_4$ takes place on the intersection $\Delta$ if and only if the following four conditions are satisfied
\begin{enumerate}
\item $(f_1^{\bot_{\Gamma_a}}<0) \wedge (f_2^{\bot_{\Gamma_a}}>0)$ which happens when $|\left(A-F_{c_2}\right)|<(\frac{m+M_1}{M_1})\cdot F_{c_1}$.

\item $(f_1^{\bot_{\Gamma_b}}<0) \wedge (f_4^{\bot_{\Gamma_b}}>0)$ which happens when $|\left(A-F_{c_1}\right)|<(\frac{m+M_2}{M_2})\cdot F_{c_2}$.

\item $(f_2^{\bot_{\Gamma_b}}<0) \wedge (f_3^{\bot_{\Gamma_b}}>0)$ which happens when $|\left(A+F_{c_1}\right)|<(\frac{m+M_2}{M_2})\cdot F_{c_2}$.

\item $(f_4^{\bot_{\Gamma_a}}<0) \wedge (f_3^{\bot_{\Gamma_a}}>0)$ which happens when $|\left(A+F_{c_2}\right)|<(\frac{m+M_1}{M_1})\cdot F_{c_1}$.
\end{enumerate}
The resulting sliding vector field $f_{s_\Delta}$ is$:$
\begin{equation}
f_{s_\Delta}=\left \{ {\begin{array}{cc}  v_m\ \ \ \ \ \ \ \ \ \ \ \ \ \ \ \ \ \ \ \ \ \  \\ \frac{1}{m+M_1+M_2} (u-kx_m)\\  v_{M_1}\ \ \ \ \ \ \ \ \ \ \ \ \ \ \ \ \ \ \ \ \ \ \\ \frac{1}{m+M_1+M_2} (u-kx_m) \\ v_{M_2} \ \ \ \ \ \ \ \ \ \ \ \ \ \ \ \ \ \ \ \ \ \ \\ \frac{1}{m+M_1+M_2} (u-kx_m)\end{array} } \right \}
\label{simpleequation}\hspace{1.0cm}
\end{equation}
\end {itemize}
\section{Numerical Approach$:$}
\noindent The algorithm has to detect$:$ (i) when a switching manifold of different dimension is reached, and (ii) whether the trajectory stays on or leaves the switching manifold. This has to be decided depending on the gradients of the continuous time behavior in the neighborhood of the current state. Let $\Gamma=\bigcup_{j=1}^p\Gamma_j$ be the entire discontinuity region.
The algorithm has to perform the following tasks$:$ (1) Robust integration outside the entire discontinuity region $\Gamma$; (2) Accurate detection and location of the switch points $x_m$ when the solution trajectory reaches $\Gamma$, this includes switching intersections for lower dimensions; (3) Check at $x_m\in\Gamma$ for the existence of a regular switching (transversality) as well as for the existence of an attractive chattering behavior (sliding motion); (4) Integration with a sliding mode on $\Gamma$; (5) Decision of whether or not we should leave the sliding region.
We would like to mention that our contribution is \textbf{independent} of using particular integration schemes. For discretizing the differential equations outside the discontinuity region $\Gamma$ any explicit integration scheme can be used. In this work we have used the mid-point rule of explicit $2^{nd}$ order Runge-Kutta integration scheme \citep{Butcher, Cellier, Munz}.
At the end of each integration step we check  whether the discontinuity region $\Gamma$ has been reached or not. We distinuiush the following cases$:$
\begin{enumerate}
\item If $\langle\gamma_j(x_{i})\cdot\gamma_j(x_{i+1})\rangle>0$ for all $j=1,2,...,p$ where $p$ is the number of the intersected $\mathbb{R}^{(n-1)}$ switching manifolds, then we continue integrating the system with the same flow map vector $f_q$.
\item If there exist $j\in\{1,2,...,p\}$ such that $\langle\gamma_j(x_{i})\cdot\gamma_j(x_{i+1})\rangle<0$, this indicates a zero crossing in the time interval $(t_i, t_i+\Delta t_i)$. In this case we have a continuous smooth switching function $\gamma_j(x_{i+1}(\sigma))$ taking opposed signs at $\sigma=0$ and $\sigma=\Delta T_i$ and therefore there exist a zero at $\sigma_m\in(0, \Delta T_i)$ which defines the switch point $x_m=x_{i+1}(\sigma_m)\in\Gamma_j$.
\item The case in which we have $\langle\gamma_j(x_{i})\cdot\gamma_j(x_{i+1})\rangle<0 \wedge \gamma_j(\sigma_m)=0$ for all $j=1,2,...,k$ where $k\leq p$ and $\sigma_m\in(0, \Delta T_i)$ indicates that the solution trajectory has reached the intersection of $k\leq p$ of transvrsally intersected $\mathbb{R}^{(n-1)}$ hyper switching manifolds $\Gamma_j$.
\end{enumerate}
In the last two cases, secant approach \citep{Myron,Kaw} is employed to find the root $\sigma_m$ of the switching function $\gamma_j(\sigma)=\gamma(x_{i+1}(\sigma))$.
Once the switch point $x_m$ has been located, the algorithm checks whether the switch point $x_m$ is a transversality point (leads to cross $\Gamma$) or a chattering equilibrium (leads to slide on $\Gamma$) by checking the sign of the two matrices $\Psi$ and $F$ (Lemma 1). The proposed convexification (Equation 40, 44-48) is used to approximate the system dynamics when sliding on $\Gamma$. For the integration during the sliding any implicit integration scheme can be used. In this work we have used the midpoint rule of implicit Bathe time integration scheme \citep{Bathe}.
To avoid any situation in which the intermediate stages values don't lie exactly on the sliding surface we use a projection formulation for convex sets together with semi-smooth Newton methods to project back to the sliding region the approximated stage values $x_{i+\frac{1}{2}}$ and $x_{i+1}$ at $t_i+\frac{\Delta t_i}{2}$ and $t_i+\Delta t_i$, respectively.
The objective of the projection formulation is to stabilize the solution during the sliding on $\Gamma$. Consider a point $x^*$ close to the hyper switching manifold $\Gamma$, each one integration step of the projected Bathe scheme on the sliding manifold $\Gamma_s$ then is given as
\begin {enumerate}
\item $x^*_{i+\frac{1}{2}}=x_i+\frac{\Delta t_i}{4} \left(f_s(x_i)+f_s(x_{i+\frac{1}{2}})\right)\;$
\item $x_{i+\frac{1}{2}}=proj(x^*_{i+\frac{1}{2}})\;$
\item $x^*_{i+1}=-\frac{1}{3}x_i+\frac{4}{3}x_{i+\frac{1}{2}}+\frac{\Delta t_i}{3}f_s(x_{i+1})\;$
\item $x_{i+1}=proj(x^*_{i+1})\;$
\end {enumerate}
The projection point $x\in\Gamma$ to $x^*$$:$ $x=P(x^*)$ is given by the solution of the projection function$:$
\begin{equation}
P:\underset{x\in\Gamma_s}{min} \ u(x): \ \ u(x)=\left(\frac{1}{2}(x^*-x)^T(x^*-x) \right)
\label{simpleequation}
\end{equation}
where $\Gamma_s$ is the sliding manifold. To find the projected value $x$ we introduce a Lagrange multiplier $\lambda$ and we need to find the root of the equation system
\begin{equation}
G(x,\lambda)= \left ( {\begin{array}{cc}
\frac{\partial u(x)}{\partial x}+\lambda \frac{\partial \gamma(x)}{\partial x}\\
\gamma(x)
\end{array} } \right )
\label{simpleequation}\hspace{2cm}
\end{equation}
We use Newton iteration, by imposing the condition $G(x_{k+1},\lambda_{k+1})=0$ and expand $G(x_{k+1},\lambda_{k+1})$ in Taylor series neglecting the terms of order equal or greater than 2.
\begin{equation}
0=G(x_k,\lambda_k)+\left ( {\begin{array}{cc}
x_{k+1}-x_k\\
\lambda_{k+1}-\lambda_k
\end{array} } \right )\cdot
G^{\prime}(x_k,\lambda_k)
\label{simpleequation}\hspace{0.0cm}
\end{equation}
for $k\geq0 $ so that
\begin{equation}
G^{\prime}(x_k,\lambda_k)\cdot \left ( {\begin{array}{cc}
x_{k+1}-x_k\\
\lambda_{k+1}-\lambda_k
\end{array} } \right )=-G(x_k,\lambda_k)
\label{simpleequation}\hspace{0.3cm}
\end{equation}
\begin{equation}
G^{\prime}(x,\lambda)=\left [ {\begin{array}{cc}
I+\lambda \frac{\partial}{\partial x}(\frac{\partial \gamma(x)}{\partial x})\ \ \ \ \ \frac{\partial \gamma(x)}{\partial x}\\
\ \ \ \ \ \left(\frac{\partial \gamma(x)}{\partial x}\right)^T \ \ \ \ \ \ \ \ \ \ \ \ 0
\end{array} } \right ]
\label{simpleequation}\hspace{0.7cm}
\end{equation}
At the end of each integration step from $t_i$ to $t_i+\Delta t_i$ while we integrate with the sliding vector $f_s$, we keep checking whether we have to keep sliding on $\Gamma_s$ or leaving it, by checking the value of $\kappa_i$ in (45).
\section{Simulation Results$:$}
\noindent As shown in Figure 3, the system in Case Study 1 was simulated for $m =M_1=M_2= 1 [kg]$, $k = 0.88 [N\cdot m^{-1}]$, $F_{c_1} = 0.01996 [N]$ $F_{c_2} = 0.062 [N]$, and $x_0=[0.8295\ 0.8491\ 0.3725\ 0.5932\ 0.8726\ 0.9335]^T$. The force $u$ was simulated as a sine wave of frequency of $\omega =0.073 [rad/sec]$. The relative and absolute error tolerance used in the approximation were adjusted to $ATOL = RTOL = 10^{-8}$.
\begin{figure}[H]
	\centering
	\subfigure[]
	{
		{\epsfig{file = 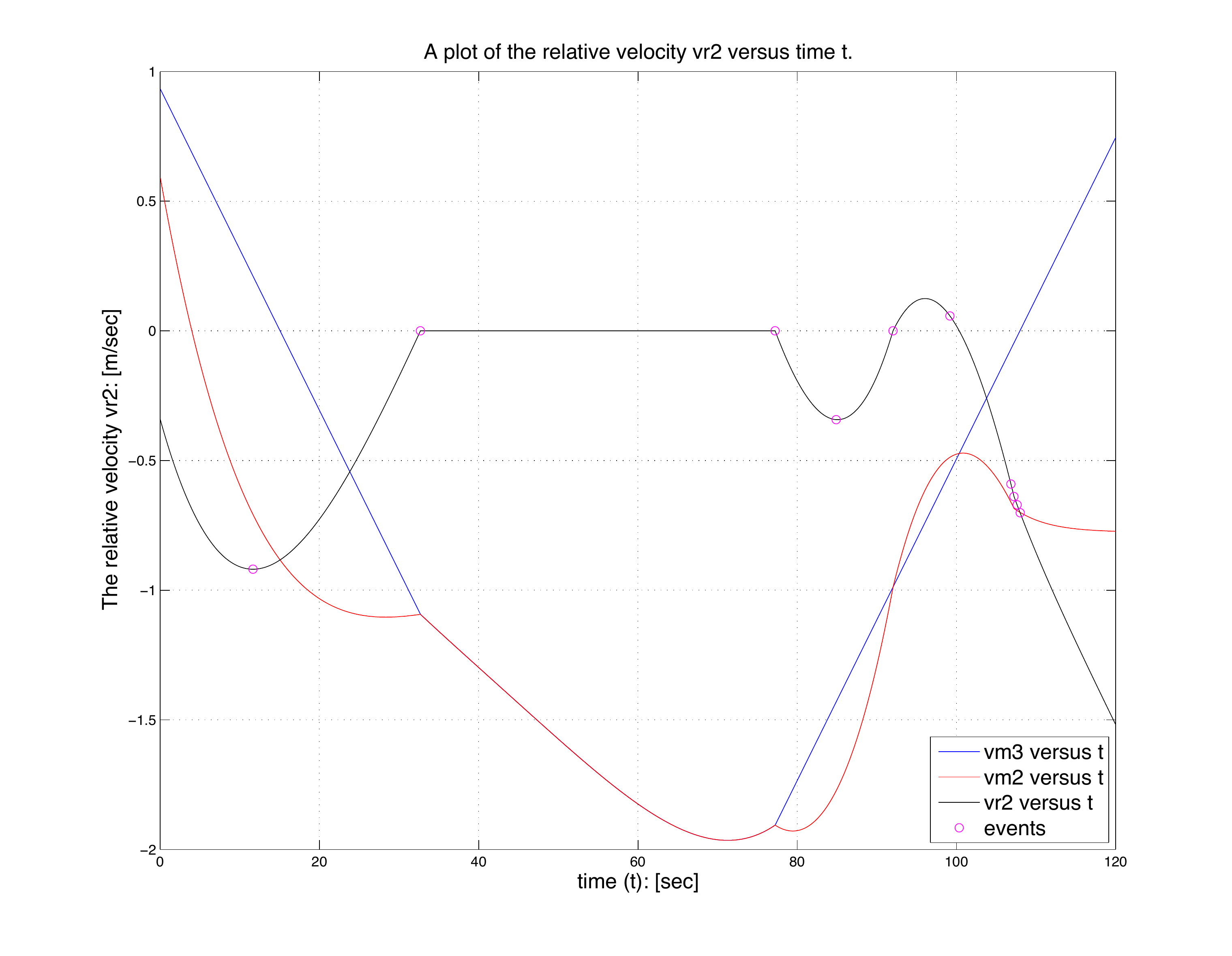, width = 9cm}}
		\label{fig:first_sub}
	}
	\\
	\subfigure[]
	{
		{\epsfig{file = 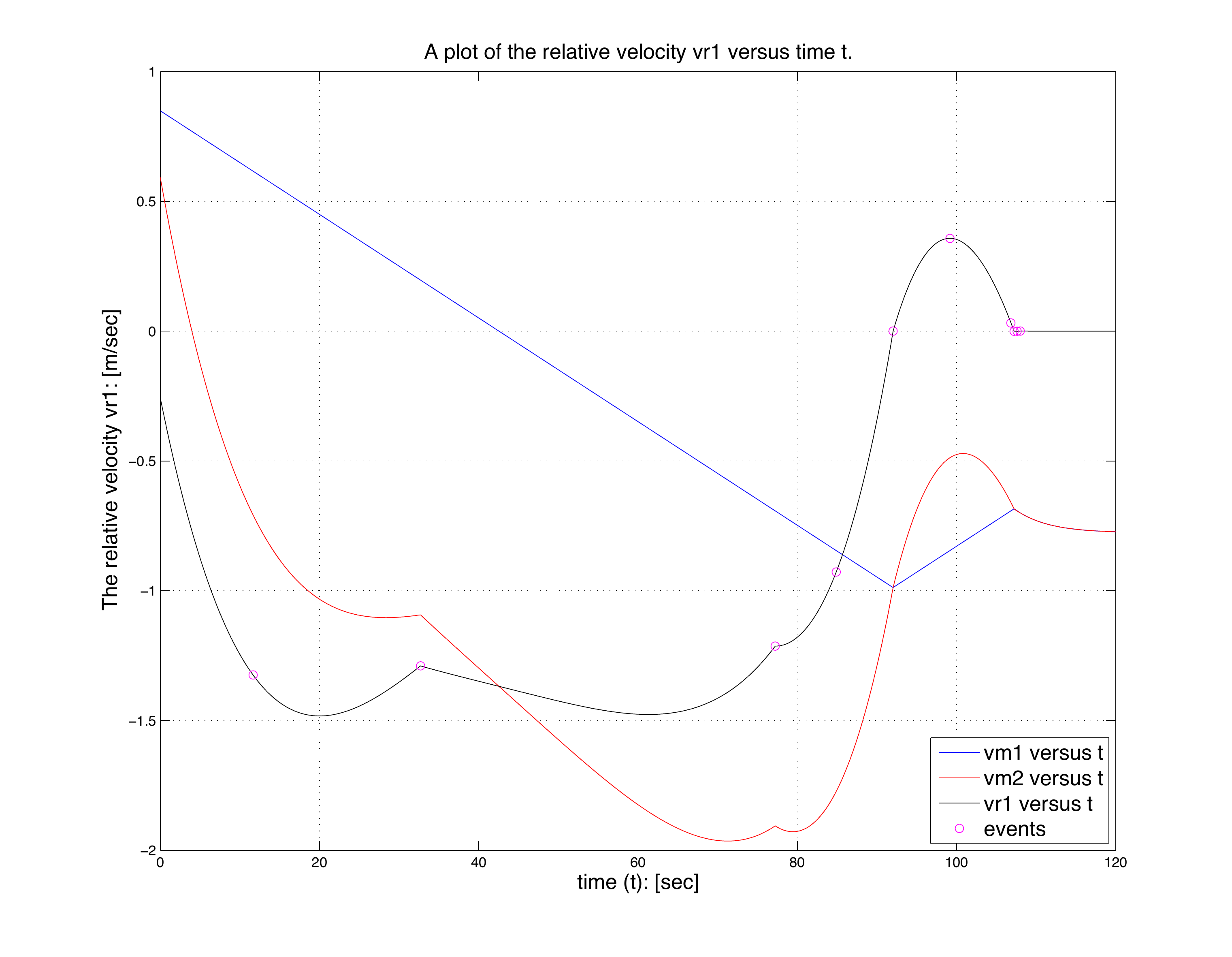, width = 9cm}}
		\label{fig:third_sub}
	}
	\caption{The sliding mode simulation of Example 2.4 with $m =M_1=M_2= 1 [kg]$, $k = 0.88 [N\cdot m^{-1}]$, $F_{c_1} = 0.01996 [N]$ $F_{c_2} = 0.062 [N]$, and $x_0=[0.8295\ 0.8491\ 0.3725\ 0.5932\ 0.8726\ 0.9335]^T$. In (a): The time evolution of the relative velocity $v_{r_2}$. In (b): The time evolution of the velocities $v_{r_1}$.}
	\label{fig:sample_subfigures}
\end{figure}
\noindent The sliding bifurcations depend on the effect of the external force $u$ and the level of Coulomb friction. We can observe that, at the beginning of the simulation, the execution of the hybrid automaton starts in the discrete state $q_3=Slip_1^-Slip_2^-$, the system starts in a slip phase for the two frictional interfaces in the system; since the effect of the external force applied tangentially on the interface is lower than the level of Coulomb friction. At the time instant $t=32.69$ $sec$, two masses $m$ and $M_2$ stick together and the solution trajectory start a sliding motion on the switching manifold $\Gamma_{b_2}$ (see Figure 3 (a)) ensuring a chattering path avoidance between the discrete states $q_3=Slip_1^-Slip_2^-$ and $q_2=Slip_1^-Slip_2^+$. A smooth exit from the sliding motion on $\Gamma_{b_2}$ into $q_3$ was detected at the time instant $t=77.23$ $sec$. A regular switching (transversality) from the discrete state $q_3=Slip_1^-Slip_2^-$ to the discrete state  $q_1=Slip_1^+Slip_2^+$ at the intersection $\Delta=\Gamma_a\cap\Gamma_b$ was detected at $t=92.04$ $sec$. At $t=108$ $sec$, two masses $m$ and $M_1$ stick together and the solution trajectory start a sliding motion on the switching manifold $\Gamma_{a_2}$  ensuring a chattering path avoidance between the discrete states $q_4=Slip_1^+Slip_2^-$ and $q_3=Slip_1^-Slip_2^-$. During a simulation time of $120$ seconds, $5$ mode switches have been recorded. \\ \\
\noindent \textbf{Case Study 2}$:$\\
Consider a mechanical system as depicted in Figure 4. The system consists of three blocks of masses $m_1$, $m_2$, and $m_3$ on a moving belt with constant velocity $v_d$. The three blocks are connected along a line by two linear springs of stiffness $k_{12}$ and $k_{23}$, and connected to a fix support by to linear springs of stiffness $k_{1}$, $k_2$, and $k_3$, respectively.
\begin{figure}[H]
	\centering
	{\epsfig{file = 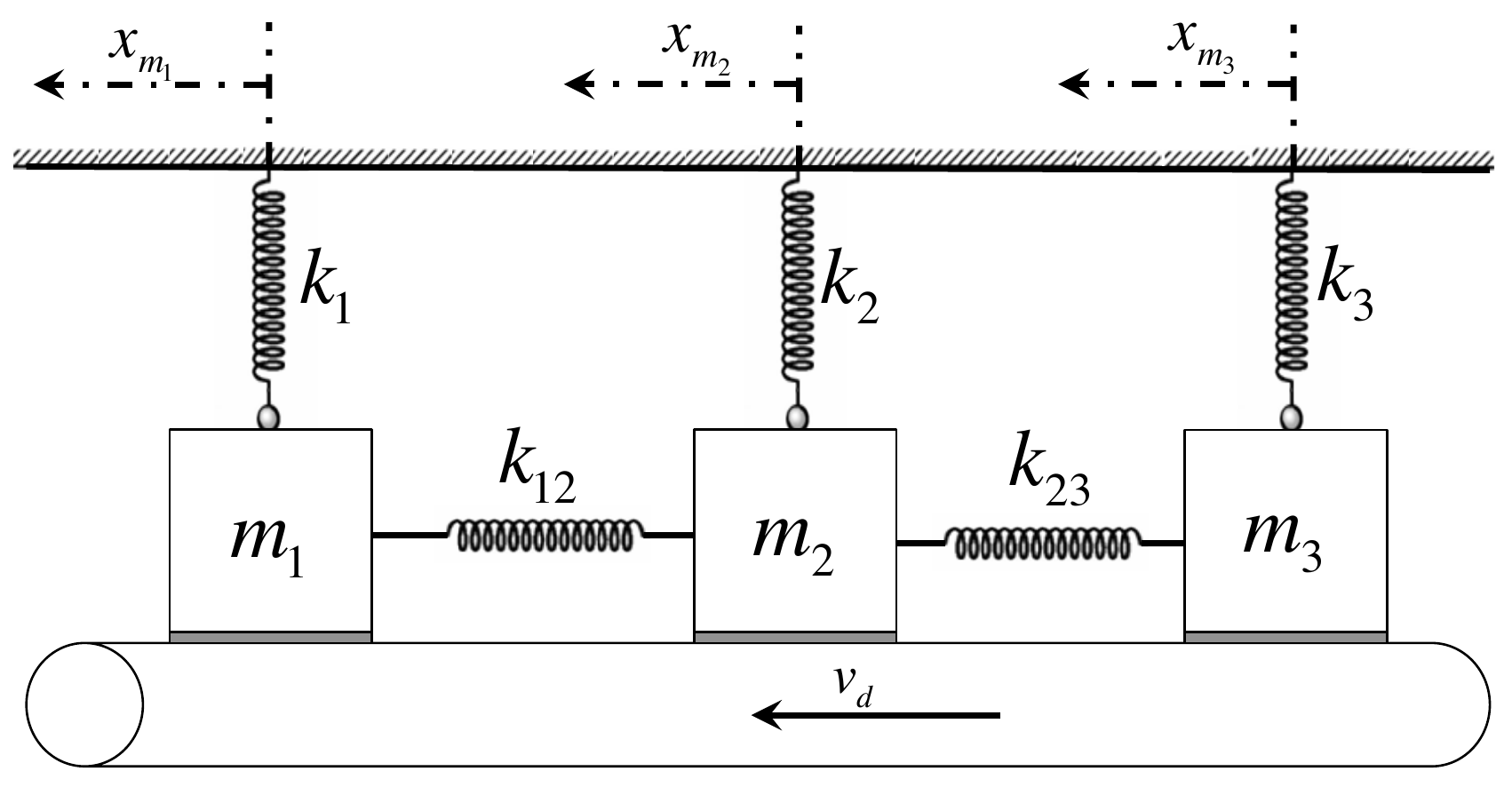, width = 7.0cm}}
	\caption{The schematic of the studied Mechanical stick-slip system with 3 blocks.}
	\label{fig:example1}
\end{figure}
\noindent The equations of the system model in $\mathbb{R}^6$ are given by
\begin{equation}
f(x):  \left \{ {\begin{array}{cc}
	\dot x_{m_1}(t)=v_{m_1}(t) \ \ \ \ \ \ \ \ \ \ \ \ \  \ \  \ \ \ \ \ \ \ \ \ \ \ \ \ \\
	\dot v_{m_1}(t)=\frac{1}{m_1}(u_1-k_1x_{m_1}(t)-\mathcal{F}_1)\\
	\dot x_{m_2}(t)=v_{m_2}(t) \ \ \ \ \ \ \ \ \ \ \ \ \  \ \  \ \ \ \ \ \ \ \ \ \ \ \ \ \ \\
	\dot v_{m_2}(t)=\frac{1}{m_2}(u_2-k_2x_{m_2}(t)-\mathcal{F}_2)\\
	\dot x_{m_3}(t)=v_{m_3}(t) \ \ \ \ \ \ \ \ \ \ \ \ \  \ \  \ \ \ \ \ \ \ \ \ \ \ \ \ \ \\
	\dot v_{m_3}(t)=\frac{1}{m_3}(u_3-k_3x_{m_3}(t)-\mathcal{F}_3)\\
	\end{array} } \right \}
\label{simpleequation}
\end{equation}
\begin{equation}
u_1=k_{12}(x_{m_2}-x_{m_1})+k_{13}(x_{m_3}-x_{m_1})
\label{simpleequation}
\end{equation}
\begin{equation}
u_2=k_{12}(x_{m_1}-x_{m_2})+k_{23}(x_{m_3}-x_{m_2})
\label{simpleequation}
\end{equation}
\begin{equation}
u_3=k_{13}(x_{m_1}-x_{m_3})+k_{23}(x_{m_2}-x_{m_3})
\label{simpleequation}
\end{equation}
We denote $x_{m_1}$, $x_{m_2}$, and $x_{m_3}$ to the position of the masses $m_1$, $m_2$, and $m_3$ respectively, $\mathcal{F}_1$, $\mathcal{F}_2$, and $\mathcal{F}_3$ to the tangential contact force on the frictional interfaces between the moving belt and the three masses $m_1$, $m_2$, and $m_3$ respectively. The functional relationship between the friction force and the relative velocity on the three frictional interfaces between the moving belt and the blocks $m_1$, $m_2$, and $m_3$ is given by
\begin{equation}
\mathcal{F}_1=\left \{ {\begin{array}{cc}
\color{white}-\color{black}F_{c_1} \ \ for \ \ v_{m_1}>v_d\\
-F_{c_1} \ \ for \ \ v_{m_1}<v_d\\
\end{array} } \right \}
\label{simpleequation}
\end{equation}
\begin{equation}
\mathcal{F}_2=\left \{ {\begin{array}{cc}
\color{white}-\color{black}F_{c_2} \ \ for \ \ v_{m_2}>v_d\\
-F_{c_2} \ \ for \ \ v_{m_2}<v_d\\
\end{array} } \right \}
\label{simpleequation}
\end{equation}
\begin{equation}
\mathcal{F}_3=\left \{ {\begin{array}{cc}
\color{white}-\color{black}F_{c_3} \ \ for \ \ v_{m_3}>v_d\\
-F_{c_3} \ \ for \ \ v_{m_3}<v_d\\
\end{array} } \right \}
\label{simpleequation}
\end{equation}
where $v_{m_1}$, $v_{m_2}$, and $v_{m_3}$ are the velocities of $m_1$,$m_2$, and $m_3$ respectively. $F_{c_1}$, $F_{c_2}$, and $F_{c_3}$ are the levels of the coulomb friction.
Physically, for each one of the three frictional interfaces in the system, as long as the force acting at the interface (call it $\rho_j$) does not exceed the Coulomb friction level $F_{c_j}$, the moving belt and the mass $m_j$ move together with $v_{r_j} = 0$ where $j=1,2,3$. As soon as $\rho_j$ exceeds the Coulomb friction level, the mass $m_j$ slips over the belt with $v_{r_j}\not= 0$. The slip is said to be positive ($Slip_j^+$) with a friction force $\mathcal{F}_j = +F_{c_j}$, i.e. the mass $m_j$ slips on the belt, if $v_j > v_{d}$. The slip motion is said to be negative ($Slip_j^-$) with a friction force $\mathcal{F} = -F_{c_j}$, if $v_j < v_d$.
Obviously the system is discontinuous on three hyper switching manifolds $\Gamma_a$, $\Gamma_b$, and $\Gamma_c$ characterized as zero sets of the switching functions $\gamma_a(x)$, $\gamma_b(x)$, and $\gamma_c(x)$ respectively
\begin{equation}
\Gamma_a=\{x\in\mathbb{R}^n: \gamma_a(x)\models v_{m_1}(t)=v_d(t)\}
\label{simpleequation}
\end{equation}
\begin{equation}
\Gamma_b=\{x\in\mathbb{R}^n: \gamma_b(x)\models v_{m_2}(t)=v_d(t)\}
\label{simpleequation}
\end{equation}
\begin{equation}
\Gamma_c=\{x\in\mathbb{R}^n: \gamma_c(x)\models v_{m_3}(t)=v_d(t)\}
\label{simpleequation}
\end{equation}
The system was simulated for the data set$:$ $m_1 =m_2=m_3= 1 [kg]$, $F_{c_1} = 0.14 [N]$, $F_{c_2} =0.13 [N]$, $F_{c_3} =0.12 [N]$, $k_1=k_2=k_3=k_{12}=k_{13}=k_{23}=0.01 [N\cdot m^{-1}]$, $v_d=0.5 m/sec$ and $x_0=[4.7799\ 0.2797\ 4.0038\ 1.7144\ 1.2922\ 4.1263]^T$. The external force $u$ was simulated as a sine wave of frequency of $\omega =0.24 [rad/sec]$. The relative and absolute error tolerance used in the approximation were adjusted to $ATOL = RTOL = 10^{-8}$.\\
The simulation results for simulation time adjusted to $100$ seconds are shown in Figure 5. At the beginning of the simulation, the execution of the hybrid automaton starts in the discrete state $q_1=Slip_1^+Slip_2^+Slip_3^+$ (i.e. the system starts in a slip phase), since the effect of the external force applied tangentially on the frictional interface is greater than the level of Coulomb friction, for the three frictional interfaces in the system.
Both $m_1$ and $m_2$ stick with the moving belt in the time interval from $t=74.84 sec$ to $t=89.07 sec$ and the solution trajectory start a sliding motion on the intersection $\Gamma_a\cap\Gamma_b=\{x\in\mathbb{R}^6: v_{m_1}=v_{m_2}=v_d\}$. At the time instant $t=76.69 sec$ the solution passes through the intersection of the three transversally intersected switching manifolds $\Gamma_a\cap\Gamma_b\cap\Gamma_c=\{x\in\mathbb{R}^6: v_{m_1}=v_{m_2}=v_{m_3}=v_d\}$ switching from a sliding on the intersection $\Gamma_a\cap\Gamma_b$ in the positive direction of the switching manifold $\Gamma_c$ into sliding on the intersection $\Gamma_a\cap\Gamma_b$ in the negative direction of $\Gamma_c$ providing a chattering bath avoidance between the flow sets associated to these intersections. During a simulation time of $100$ seconds, $16$ mode switches as well as $12$ tangential crossings have been recorded.
\begin{figure}[H]
    \centering
    \subfigure[]
    {
   {\epsfig{file = 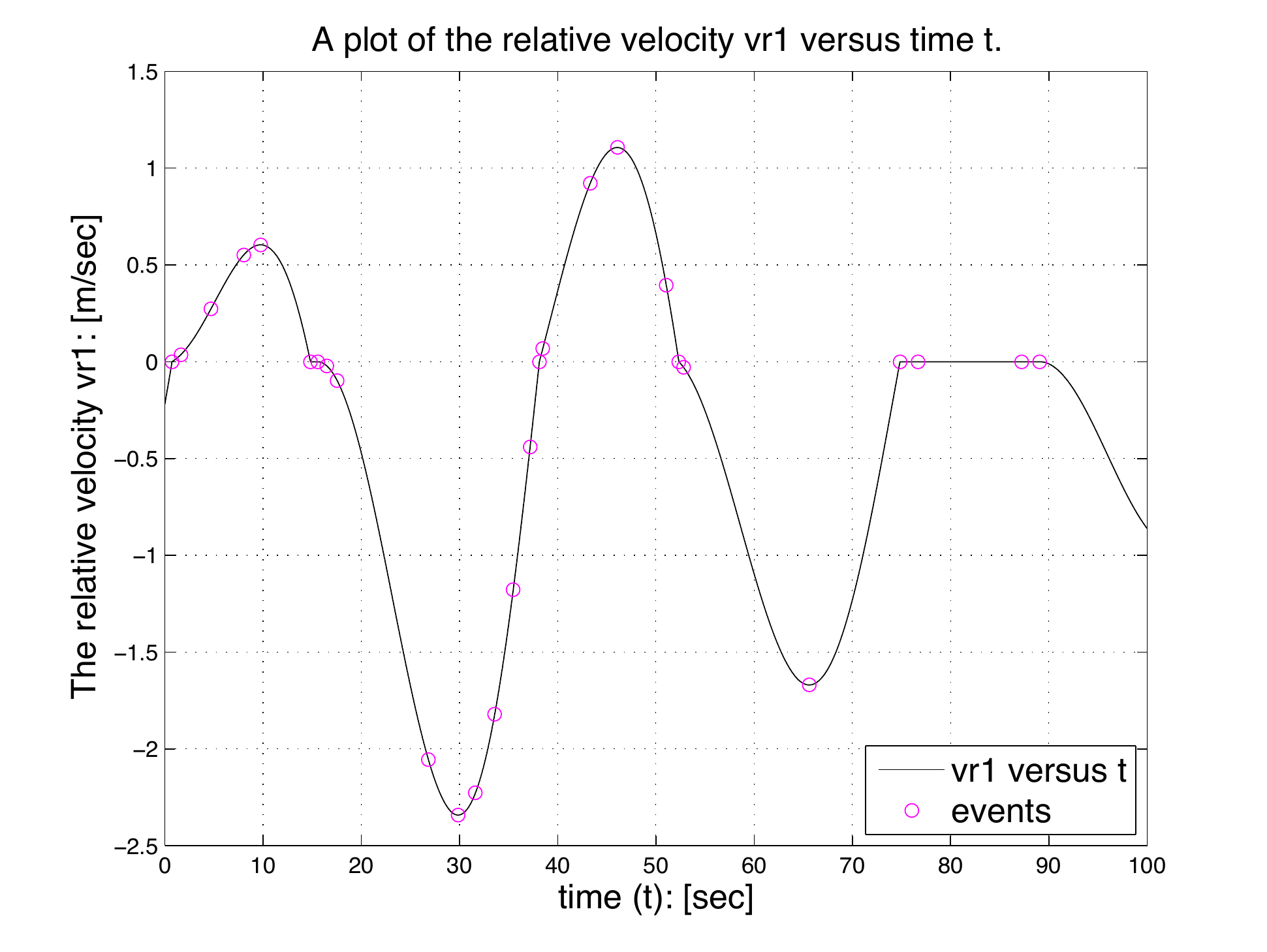, width = 9cm}}
        \label{fig:first_sub}
    }
       \\
     \subfigure[]
    {
   {\epsfig{file = 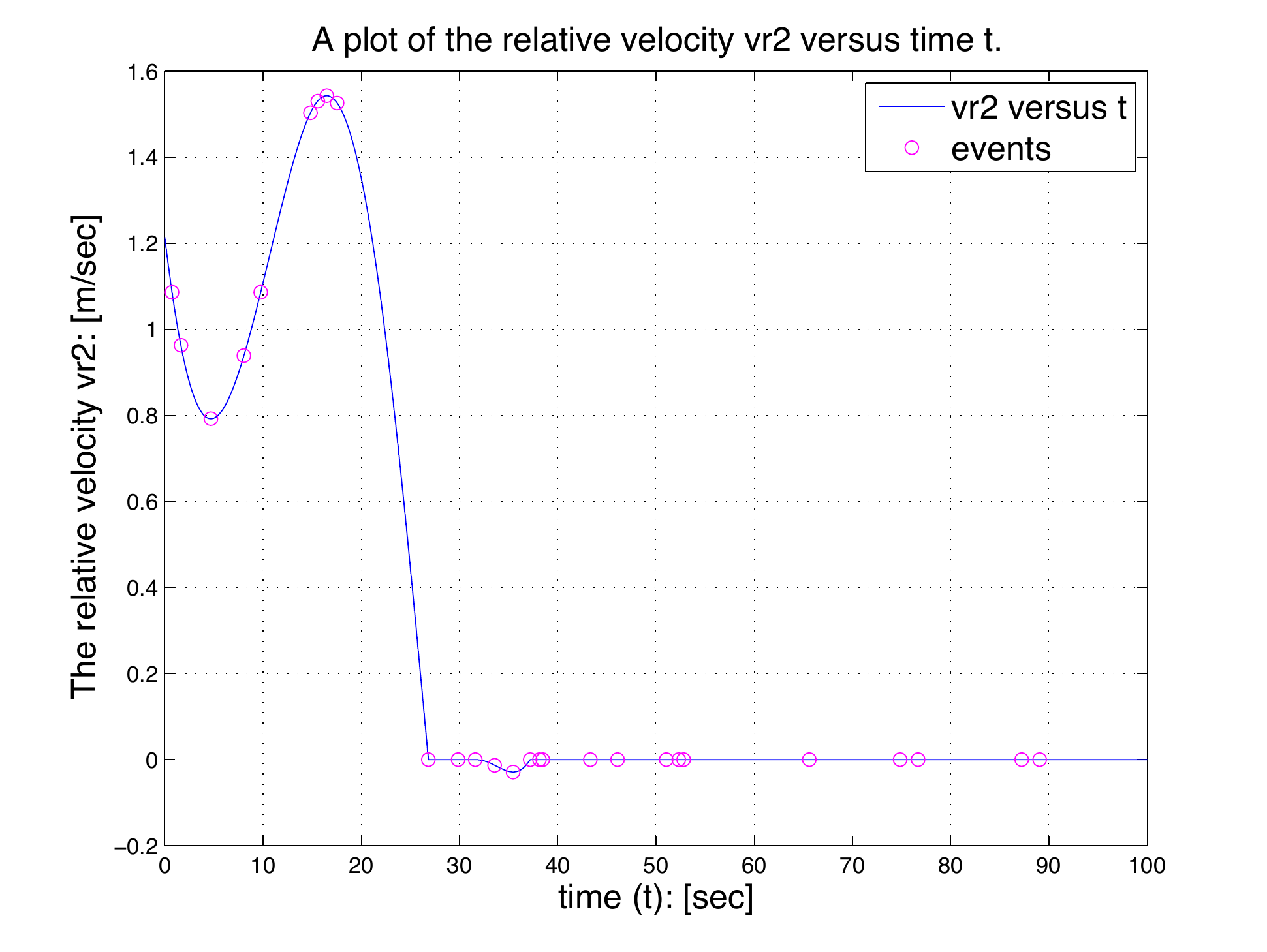, width = 9cm}}
        \label{fig:third_sub}
    }
    \caption{The sliding mode simulation of Case Study 2 with $m_1 =m_2=m_3= 1 [kg]$, $k_1=k_2=k_3=k_{12}=k_{13}=k_{23}=0.01 [N\cdot m^{-1}]$, $F_{c_1} = 0.14 [N]$ $F_{c_2} =0.13 [N]$, $F_{c_3} =0.12 [N]$, $v_d=0.5 m/sec$, and $x_0=[4.7799\ 0.2797\ 4.0038\ 1.7144\ 1.2922\ 4.1263]^T$. In (a): The time evolution of the relative velocity $v_{r_1}$. In (b): The time evolution of the relative velocity $v_{r_2}$.}
    \label{fig:sample_subfigures}
\end{figure}
\section{Conclusions}
\noindent In this paper we presented a new approach for the robust and stable numerical simulation of hybrid systems with chattering behavior, characterized  either by infinite fast control actions or by discontinuous laws of physics. We have proposed a technique to detect chattering behavior “on the fly” in real-time simulation. Necessary and sufficient conditions driving to chattering behavior are explicitly introduced, and the treatment of chattering behavior during the numerical simulation using so-called sliding mode simulation is studied. A simulation algorithm is proposed that guarantees a robust treatment of chattering behavior. Its purpose is to organize mode switching and to allow sliding mode simulation each time the necessary and sufficient condition for chattering is satisfied. A novel computational framework which treats automatically any non-smoothness in the trajectory of the state variables during the regularization of the chattering path by a smooth correction after each integration time-step was also provided. The main objective of the computational algorithm is to switch between the transversality modes and the sliding modes simulation automatically as well as integrating each particular state appropriately and localize the structural changes in the system in an accurate way. Our approach is based on mixing compile-time transformations of hybrid programs (generating what is necessary to compute the smooth equivalent dynamics), the decision at run-time of the necessary and sufficient conditions for entering and exiting a sliding mode, and the computation, at run-time, of the smooth equivalent dynamics.
We have shown by a special hierarchical application of convex combinations that unique solutions can be found in general cases when the switching function takes the form of finitely many intersecting manifolds so that an efficient numerical treatment of the sliding motion constrained on the entire discontinuity region (including the switching intersection) is guaranteed.  \\
Finally, the simulation results - reported here on a set of representative case studies - showed that our approach is efficient and precise enough to provide a chattering bath avoidance, to perform a special numerical treatment of the constrained motion along the discontinuity surface, as well as its robustness in achieving an accurate detection and localization of the switch points for both entering to- and exiting from- the sliding region for the case of a single manifold of discointuniuity as well as the case of switching intersection.

\section*{Acknowledgements}
\noindent This work was supported by the ITEA2/MODRIO project under contract N\textsuperscript{o} 6892, and the ARED grant of the Brittany Regional Council.\\ \\

\small
\bibliographystyle{plainnat}
\bibliography{References}
\normalsize

\end{document}